\documentclass[pra,showpacs]{revtex4}

\usepackage{graphicx}
\usepackage{bm}
\usepackage[english]{babel}
\usepackage{dcolumn}
\usepackage{amssymb}
\usepackage{amsmath}

\begin{document}

\title{Bound cyclic systems with the envelope theory}

\author{Claude \surname{Semay}}
\email[E-mail: ]{claude.semay@umons.ac.be}
\affiliation{Service de Physique Nucl\'{e}aire et Subnucl\'{e}aire,
Universit\'{e} de Mons,
UMONS Research Institute for Complex Systems,
Place du Parc 20, 7000 Mons, Belgium}
\author{Fabien \surname{Buisseret}}
\email[E-mail: ]{fabien.buisseret@umons.ac.be}
\affiliation{Service de Physique Nucl\'{e}aire et Subnucl\'{e}aire,
Universit\'{e} de Mons,
UMONS Research Institute for Complex Systems,
Place du Parc 20, 7000 Mons, Belgium;\\ 
Haute \' Ecole Louvain en Hainaut (HELHa), Chauss\'ee de Binche 159, 7000 Mons, Belgium}
\date{\today}

\begin{abstract}
Approximate but reliable solutions of a quantum system with $N$ identical particles can be easily computed with the envelope theory, also known as the auxiliary field method. This technique has been developed for Hamiltonians with arbitrary kinematics and one- or two-body potentials. It is adapted here for cyclic systems with $N$ identical particles, that is to say systems in which a particle $i$ has only an interaction with particles $i-1$ and $i+1$ (with $N+1\equiv 1$). 
%03.65.Ge Solutions of wave equations: bound states
\end{abstract}

\maketitle

\section{Introduction}
\label{sec:intro}

Several methods are available to solve the $N$-body quantum problems. Among the most accurate, one can find the Gaussian expansion \cite{suzu98}, the oscillator bases \cite{zouz86},  the Faddeev formalism \cite{silv96}, or the Lagrange-mesh method \cite{baye15}. These procedures are nevertheless quite heavy to implement and require long computation times increasing dramatically with the number of particles, especially when relativistic kinematics is chosen. So, it can be useful to consider simpler methods to obtain approximate but reliable solutions. 

The envelope theory (ET) is such a technique \cite{hall80,hall04}. It has been independently rediscovered under the name of auxiliary field method \cite{silv10} and it has been extended to treat systems with arbitrary kinematics in $D$-dimensions \cite{sema13a}. The basic idea is to replace the Hamiltonian $H$ under study by an auxiliary Hamiltonian $\tilde H$ which is solvable, the eigenvalues of $\tilde H$ being optimized to be as close as possible to those of $H$. The method is easy to implement since it reduces to find the solution of a transcendental equation. Recently, the accuracy of the ET has been tested for eigenvalues and eigenvectors by computing the ground state of various systems containing up to 10 bosons \cite{sema15a}. This comparison was possible thanks to accurate numerical results published in \cite{horn14}.
The solutions obtained by the ET can be used as tests for heavy numerical computations. Moreover, if a lower or an upper bound can be computed, the information they bring may be sufficient for some applications. In the peculiar situations where an analytical expression is obtained, the results can give valuable insights about the system, the dependence of the energy spectrum on the various parameters of the model in particular. Detailed properties of the ET have been exposed in \cite{silv11,sema15c}, to which we refer the interested reader. Its accuracy can be improved, but to the detriment of the variational character \cite{sema15b}.

\begin{figure*}[htb]
% 1513 x 369
\includegraphics[width=11.3cm,height=3cm,keepaspectratio=true]{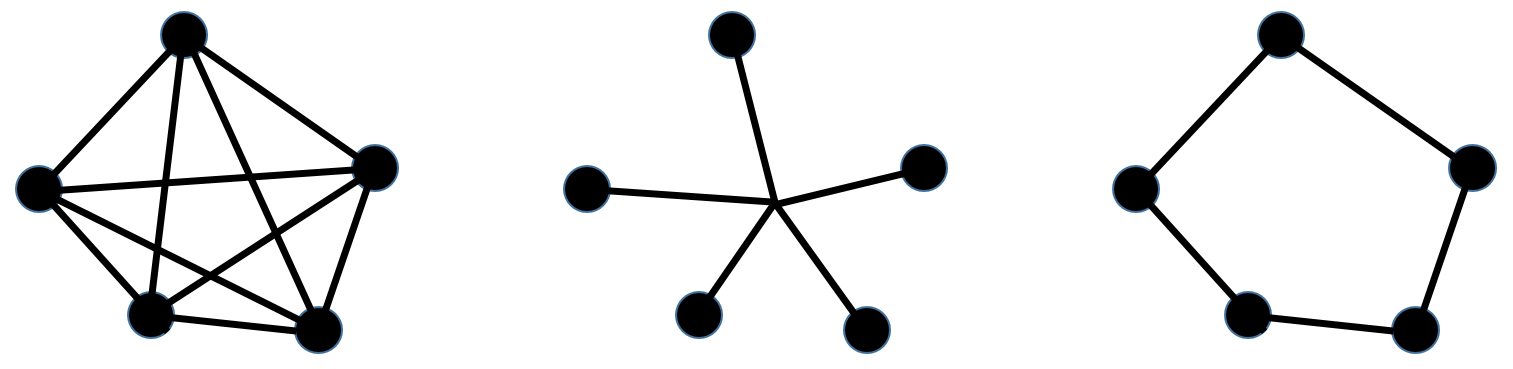}
\caption{From left to right, schematic representation of the interactions for 5-body systems with two-body interactions, one-body interactions (from the center of mass), and cyclic-interactions. The first two systems are described by Hamiltonians of the form (\ref{HNb}), while the last system corresponds to (\ref{Hcy}).}
\label{fig:syst} 
\end{figure*}

All the results mentioned above are obtained for systems of $N$ identical particles, with a kinetic energy $T$, interacting via the one-body $U$ and two-body $V$ interactions ($\hbar=c=1$)
\begin{equation}
\label{HNb}
H=\sum_{i=1}^N T(|\bm p_i|) + \sum_{i=1}^N U\left(|\bm r_i - \bm R|\right) + \sum_{i\le j=2}^N V\left(|\bm r_i - \bm r_j|\right),
\end{equation}
where $\sum_{i=1}^N \bm p_i = \bm 0$ and $\bm R = \frac{1}{N}\sum_{i=1}^N \bm r_i$ is the center of mass position. $\bm p_i$ and $\bm r_i$ are the momentum and position of particle $i$. The purpose of this work is to adapt the ET to treat cyclic systems of $N$ identical particles, for a Hamiltonian given by 
\begin{equation}
\label{Hcy}
H=\sum_{i=1}^N T(|\bm p_i|) + \sum_{i=1}^N W\left(|\bm r_i - \bm r_{i+1}|\right) ,
\end{equation}
with $r_{N+1}\equiv r_1$. Systems described by Hamiltonians (\ref{HNb}) and (\ref{Hcy}) are illustrated for a 5-body system in Fig.~\ref{fig:syst}. Cyclic systems appears obviously in organic chemistry \cite{zimm75}, but they could also play a central role in the phenomenology of glueballs \cite{iwas03,meye05,buis09}. 

In Sec.~\ref{sec:hoi}, the exact solution for the non relativistic cyclic system of $N$ identical harmonic oscillators is given. The ET treatment, based on this solution, for general cyclic Hamiltonians is developed in Sec.~\ref{sec:gi}, and some examples are computed in Sec.~\ref{sec:ex}. Concluding remarks are given in the last section.

\section{Harmonic oscillator interaction}
\label{sec:hoi}

The Hamiltonian for the non relativistic cyclic system of $N$ identical harmonic oscillators is the basis of the forthcoming application of the ET. It is given by 
\begin{equation}
\label{Hcyo}
H=\frac{1}{2 m}\sum_{i=1}^N \bm p_i^2 + \frac{m\, \omega^2}{2}\sum_{i=1}^N  \left(\bm r_i - \bm r_{i+1}\right)^2, 
\end{equation}
with $r_{N+1}\equiv r_1$. The summation in the potential can be written in matrix form as
\begin{equation}
\label{quad}
\sum_{i=1}^N  \left(\bm r_i - \bm r_{i+1}\right)^2 = \bm r\, M\, \bm r^t =
\sum_{i=1}^N  \lambda_i\,\bm x_i^2,
\end{equation}
where $\bm r=( \bm r_1\, \bm r_2\, \ldots \bm r_N)$ and 
\begin{equation}
\label{M}
M = \begin{pmatrix}
2 & -1 & 0 & 0 & \ldots & 0 & -1\\
-1 & 2 & -1 & 0 & \ldots & 0 & 0\\
0 & -1 & 2 & -1 & \ldots & 0 & 0\\
\vdots & \vdots & \vdots & \vdots & \ddots & \vdots & \vdots\\
-1 & 0 & 0 & 0 & \ldots & -1 & 2\\\end{pmatrix}.
\end{equation}
The coefficients $\lambda_i$ are the eigenvalues of $M$. They can be computed with some algebra once it is noticed that $M$ is a circulant matrix:
\begin{equation}
\label{lambda}
\lambda_i=4 \sin^2\left( \frac{i \, \pi}{N} \right),
\end{equation}
with $i=1,\, \ldots,\, N$ ($\lambda_N=0$). The internal coordinates $\bm x_i$ are given by
\begin{equation}
\label{xi}
\bm x_i= \sum_{j=1}^N U_{i j} \, \bm r_j,
\end{equation}
where
\begin{equation}
\label{U}
U_{i j} = \frac{1}{\sqrt{N}} \left[ \cos\left( i\, j\frac{2\pi}{N} \right) +
\sin\left( i\, j\frac{2\pi}{N} \right) \right],
\end{equation}
with $U^{-1}= U$ and $\bm x_N=\sqrt{N}\,\bm R$. With the coordinates $\bm x_i$, Hamiltonian~(\ref{Hcyo}) can be written
\begin{equation}
\label{Hcyxi}
H=\frac{1}{2 N m} \bm P^2 +  \sum_{i=1}^{N-1} \left(\frac{1}{2 m} \bm q_i^2 + \frac{m\, \omega^2}{2} \lambda_i\,\bm x_i^2 \right), 
\end{equation}
where the variables $\bm q_i$ are the conjugate of the coordinates $\bm x_i$, and where $\bm P$ is the total momentum of the system. In the center of mass frame ($\bm P=\bm 0$), the solutions of this Hamiltonian are given by
\begin{equation}
\label{E}
E=\omega \sum_{i=1}^{N-1} \sqrt{\lambda_i}\, \left( 2 n_i+l_i+\frac{D}{2}  \right),
\end{equation}
with $\sqrt{\lambda_i}=2\,\sin( i \, \pi/N )$, and $D$ the dimension of the space. The corresponding eigenfunctions are 
\begin{equation}
\label{phi}
\phi = \prod_{i=1}^{N-1} \varphi_{n_i l_i}(\gamma_i\, \bm x_i),
\end{equation}
where $\varphi_{n_i l_i}(\gamma_i\, \bm x_i)$ is a harmonic oscillator wave function with the size parameter $\gamma_i=\left( m^2\omega^2 \lambda_i \right)^{1/4}$ \cite{yane94}. These solutions are similar to the ones found for the treatment of cyclic molecules by the H\"uckel method \cite{zimm75}. Interestingly, an explicitly covariant version of Hamiltonian (\ref{Hcyxi})  appears when quantizing the closed Nambu-goto string in the framework of the discretized string \cite{gers10}. In this approach, the excitations of the closed string are carried by pointlike bosonic degrees of freedom linked by pieces of string: This is very similar in nature to the cyclic systems we are studying.

The strong degeneracy of the usual $N$-body harmonic oscillator \cite{silv10} is broken by the presence of $\lambda_i$ in formula~(\ref{E}). But a symmetry persists: $\lambda_i=\lambda_{N-i}$. For $N=2$ and $3$, the potential in (\ref{Hcy}) reduces respectively to $m\, \omega^2 \left(\bm r_1 - \bm r_2\right)^2$ and $m\, \omega^2 \left(\left(\bm r_1 - \bm r_2\right)^2 + \left(\bm r_2 - \bm r_3\right)^2 + \left(\bm r_3 - \bm r_1\right)^2\right)/2$. This is equivalent to usual two-body interactions. In these two cases, it is easy to verify that (\ref{E}) is the correct solution ($\lambda_1=4$ for $N=2$, while $\lambda_1=\lambda_2=3$ for $N=3$). Let us remark that for $N=6$, $2\,\lambda_1=2\,\lambda_5=\lambda_3$ and $\lambda_2=\lambda_4$. So supplementary level degeneracy appears in this particular case. 

As a particular case of (\ref{E}) and (\ref{phi}), let us explicitly write the ground state energy and wave function. With $n_i =l_i=0$, one has
\begin{equation}
\label{E_0}
E_0=\omega\, D \cot\left(  \frac{\pi}{2 N}\right).
\end{equation}
The corresponding eigenfunction is given by
\begin{equation}
\label{phi0}
\phi_0 = \prod_{i=1}^{N-1} \varphi_{0 0}(\gamma_i\, \bm x_i) = K\, \exp\left( -\frac{m\, \omega}{2} \sum_{i=1}^{N-1} \sqrt{\lambda_i}\, \bm x_i^2 \right),
\end{equation}
where $K$ is a normalization factor. Using formulas~(\ref{lambda})-(\ref{U}), $\phi_0$ can be recast under the form
\begin{equation}
\label{phi0r}
\phi_0 = K\, \exp\left( -\frac{m\, \omega}{2} \sum_{i=1}^N \sum_{j=1}^N 
Z_{i \,j}\, \bm r_i\cdot\bm r_j \right),
\end{equation}
with 
\begin{equation}
\label{Wij}
Z_{i\, j} = \frac{2}{N} \frac{\sin(\pi/N)}{\cos(é(i-j)\pi/N)-\cos(\pi/N)}.
\end{equation}
Clearly, $\phi_0$ is invariant under cyclic permutations, since $Z_{i\, j} = Z_{i+1\, j+1}$; the matrix $Z$ is also circulant.

\section{General interaction and kinetic term}
\label{sec:gi}

The procedure to obtain the ET approximation for the Hamiltonian~(\ref{Hcy}) is very similar to the one computed in \cite{silv10}. We present here the computation for the cyclic case in the main lines. The Hamiltonian $\tilde H$ is defined by
\begin{equation}
\label{Htilde}
\tilde H=\sum_{i=1}^N \left( \frac{\bm p_i^2}{2 \mu} + \rho\, \bm s_i^2 \right)
+N \left( T(G(\mu))-\frac{G(\mu)^2}{2\, \mu} + W(J(\rho))-\rho\,J(\rho)^2 \right), 
\end{equation}
where $\bm s_i=\bm r_i - \bm r_{i+1}$. The parameters $\mu$ and $\rho$ are to be determined. The functions $G$ and $J$ are such that $G^{-1}(x)=x/T'(x)$ and $J^{-1}(x)=W'(x)/(2\,x)$. These inverse functions are assumed to be defined in a domain relevant for the physical problem \cite{silv10}. Let $\tilde E = \langle \mu,\,\rho | \tilde H | \mu,\,\rho \rangle$, where $| \mu,\,\rho \rangle$ is an eigenstate of $\tilde H$ for the parameters $\mu$ and $\rho$. If the values $\mu_0$ and $\rho_0$ are such that
\begin{equation}
\label{Cond}
\left. \frac{\partial \tilde E}{\partial \mu}\right|_{\mu_0,\rho_0} = \left. \frac{\partial \tilde E}{\partial \rho}\right|_{\mu_0,\rho_0}= 0,
\end{equation}
then, the Hellmann-Feynman theorem \cite{lich89} implies that 
\begin{eqnarray}
\label{GJa}
N\, G(\mu_0)^2 &=& \sum_{i=1}^N \langle \mu_0,\,\rho_0 |\bm p_i^2 | \mu_0,\,\rho_0 \rangle, \\
\label{GJb}
N\, J(\rho_0)^2 &=& \sum_{i=1}^N \langle \mu_0,\,\rho_0 |\bm s_i^2 | \mu_0,\,\rho_0 \rangle. 
\end{eqnarray}
Posing $G(\mu_0)=p_0$ and $J(\rho_0)=r_0$, the ET approximation $E$ for an eigenvalue of Hamiltonian~(\ref{Hcy}) is given by 
\begin{equation}
\label{EET}
E = \langle \mu_0,\,\rho_0 | \tilde H_0 | \mu_0,\,\rho_0 \rangle
= N\left( T(p_0) + W(r_0) \right),
\end{equation}
where $\tilde H_0$ is $\tilde H$ with $\mu=\mu_0$ and $\rho=\rho_0$. The generalized virial theorem \cite{luch90} applied to $\tilde H_0$ gives
\begin{equation}
\label{virial}
\sum_{i=1}^N \langle \mu_0,\,\rho_0 | \frac{\bm p_i^2}{\mu} | \mu_0,\,\rho_0 \rangle
= \sum_{i=1}^N \langle \mu_0,\,\rho_0 | 2\,\rho\, \bm s_i^2 | \mu_0,\,\rho_0 \rangle,
\end{equation}
which implies
\begin{equation}
\label{r0p0}
p_0\, T'(p_0) = r_0\, W'(r_0).
\end{equation}
The link between $p_0$ and $r_0$ can be computed thanks to the knowledge of the exact solutions of $\tilde H_0$ given by (\ref{E}). Finally, 
\begin{eqnarray}
\label{forma}
E &=& N\left( T(p_0) + W(r_0) \right), \\
\label{formb}
p_0\, T'(p_0) &=& r_0\, W'(r_0) \\
\label{formc}
r_0\, p_0 &=& \frac{Q}{N}, 
\end{eqnarray}
with 
\begin{equation}
\label{Q}
Q = 2 \sum_{i=1}^{N-1}\sin\left( \frac{i\, \pi}{N} \right)\, \left( 2 n_i+l_i+\frac{D}{2}  \right).
\end{equation}
Once the global quantum number $Q$ is fixed by (\ref{Q}), $r_0$ and $p_0$ can be determined by solving the transcendental system (\ref{formb})--(\ref{formc}). The approximation $E$ for the eigenvalue can then be computed by (\ref{forma}). As a trivial test, solution~(\ref{E}) is recovered for the Hamiltonian~(\ref{Hcy}). One of the interest of the ET method is the possible existence of upper or lower bounds. Let us define two functions $b_T$ and $b_W$ such that 
\begin{equation}
\label{hg}
T(x) = b_T(x^2) \quad  \textrm{and} \quad  W(x) = b_W(x^2).
\end{equation}
It has been shown \cite {hall80,hall04} that, if $b_T''(x)$ and $b_W''(x)$ are both concave (convex) functions, $E$ is an upper (lower) bound of the genuine eigenvalue. If the second derivative is vanishing for one of these functions, the variational character is solely ruled by the convexity of the other one. In the other cases, the solution has a priori no variational character. 

Equations~(\ref{GJa}) and (\ref{GJb}) give immediately
\begin{eqnarray}
\label{p0}
p_0^2 &=& \frac{1}{N} \sum_{i=1}^N \langle \mu_0,\,\rho_0 |\bm p_i^2 | \mu_0,\,\rho_0 \rangle, \\
\label{r0}
r_0^2 &=& \frac{1}{N} \sum_{i=1}^N \langle \mu_0,\,\rho_0 |\bm s_i^2 | \mu_0,\,\rho_0 \rangle. 
\end{eqnarray}
So, $p_0$ can be considered as the mean momentum per particle, and $r_0$ as the mean distance between two neighbouring particles, which is in agreement with (\ref{forma}). The mean total length of the cyclic system is then given by $L=N\,r_0$. An approximation for an eigenfunction is given by (\ref{phi}), with 
\begin{equation}
\label{gammai}
\gamma_i = \sqrt{\frac{2\,Q}{N}\sin\left( \frac{i\, \pi}{N} \right)}\, \frac{1}{r_0}.
\end{equation}

It can be noticed that the ground state is such that $Q=D\cot(\frac{\pi}{2N})\propto N$ at large $N$. Hence, it can be deduced from (\ref{forma})-(\ref{formc}) that $E\propto N$ at large $N$, as expected. This result is independent of the form of $T$ and $W$ but requires that these functions do not depend on $N$. 

\section{Examples}\label{sec:ex}

\subsection{Homogeneous potential and kinematics}

Equations~(\ref{forma})-(\ref{Q}) are similar to those corresponding to the Hamiltonian~(\ref{HNb}) with one-body interactions \cite{sema13a}. So, analytical solutions found for such systems \cite{silv12} are readily transposable to cyclic systems. Let us choose 
\begin{equation}\label{kindef}
T(p)=A\, p^B,\quad  A,\,B\in\mathbb{R}^+.
\end{equation}
This generic kinematics covers the nonrelativistic ($B=2$) and ultrarelativistic ($B=1$) cases. Moreover, we set  $W(r)=C\, r^F$ ($C,\,F\in\mathbb{R}$ and $C\,F>0$). Then it is found after straightforward algebra that 
\begin{equation}\label{ehom}
E=N\, C\left( \frac{B+F}{B}\right) \left(\frac{A\,B}{C\,F}\right)^{\frac{F}{B+F}} \left(\frac{Q}{N}\right)^{\frac{B\,F}{B+F}}.
\end{equation}

For instance, the semirelativistic Hamiltonian of $N$ massless particles linked by identical linear potentials, which is a very simple possible model for a glueball \cite{iwas03,meye05,buis09}, is written
\begin{equation}
\label{Hg}
H=\sum_{i=1}^N |\bm p_i| + \sum_{i=1}^N \sigma\,|\bm r_i - \bm r_{i+1}|
\end{equation}
and corresponds to $A=B=F=1$, $C=\sigma$, this last parameter being the string tension. Formula (\ref{ehom})  reads in this case
\begin{equation}
\label{Eg}
E=2\sqrt{N\, \sigma\, Q}= 2\, \sigma\, L.
\end{equation}
The fact that $E^2\propto Q$ is linked to the well-known Regge trajectories appearing in any string model of hadrons. 

\subsection{Finite range potential}

Let us assume that $W(r)=-g \, w(r)$, where $g$ is a positive constant with the dimension of an  energy and $w(r)$ a dimensionless function such that $W(r)$ supports only a finite number of bound states for a given kinematics (\ref{kindef}). For a given set of quantum numbers $\{ n_i,l_i\}$, that is to say a given value of $Q$, the critical value $g_c$ allowing the existence of a bound state with these quantum numbers can be found by imposing $E=0$ in (\ref{forma})-(\ref{Q}) \cite{sema13a}. The computation gives
\begin{equation}
\label{gc}
g_c = \frac{A}{y_0^B\, w(y_0)}\left(\frac{Q}{N}\right)^B,
\end{equation}
where $y_0$ is the solution of the equation
\begin{equation}
\label{y0}
y_0\,w'(y_0)+B\, w(y_0)=0.
\end{equation}
The variable $y_0$ is independent of $N$, $Q$, and $A$. It depends only on the type of the kinematics ($B$) and on the form of the function $w(r)$.

The ground state is allowed to exist when $g\geq \frac{A}{y_0^B\, w(y_0)} \left(\frac{D}{N}\cot(\frac{\pi}{2N})\right)^B$, that is a slightly increasing function of $N$. Chains with a larger number of particles are predicted to be less strongly bound by the ET, although the effect is quite weak: For $B=2$, the critical constant at $N=3$ is only 18\% smaller than in the $N\rightarrow \infty$ limit. This result could find a relevant application in the study of chain-like bound states in the quark-gluon plasma (see \textit{e.g.} \cite{shur06}).

\section{Concluding remarks}
\label{sec:cr}

The exact solution of the non relativistic cyclic system of $N$ identical harmonic oscillators has been used to compute, in the framework of the envelope theory \cite{hall80,hall04,silv10,sema13a}, approximate solutions for cyclic systems of $N$ identical particles with arbitrary kinematics. The approximate eigenvalues can be computed as the roots of a transcendental equation, and the corresponding approximate eigenstates are built as the product of oscillator waves functions. This method is thus very simple to implement and has been proven reasonably accurate in the case of non cyclic systems with one- or two-body forces \cite{sema15a,sema15b}. 

The purpose of this procedure is not to compete with accurate numerical methods \cite{suzu98,zouz86,silv96,baye15}, but to yield rapidly a reliable solution, which can be used for instance as tests for numerical calculations. Depending on the Hamiltonian considered, upper or lower bounds can be computed, sometimes under an analytical form. These informations can be sufficient to study the main characteristics of a cyclic system. 

\section*{Acknowledgement}
F. B. thanks J. B. Coulaud and C. de Kerchove d'Exaerde for enlightening discussions about circulant matrices.


\begin{thebibliography}{99} 

\bibitem{suzu98} Y. Suzuki and K. Varga, \textit{Stochastic Variational Approach to Quantum-Mechanical Few-Body Problems} (Springer, Berlin, 1998).
\bibitem{zouz86} S. Zouzou, B Silvestre-Brac, C. Gignoux, and J. M. Richard, Four-quark bound states, Z. Phys. \textbf{30}, 457 (1986).
\bibitem{silv96} B. Silvestre-Brac, Spectrum and static properties of heavy baryons, Few-Body Syst. \textbf{20}, 1 (1996).
\bibitem{baye15} D. Baye, The Lagrange-mesh method, Phys. Rep. \textbf{565}, 1 (2015).
\bibitem{hall80} R. L. Hall, Energy trajectories for the $N$-boson problem by the method of potential envelopes, Phys. Rev. D \textbf{22}, 2062 (1980).
\bibitem{hall04} R. L. Hall, W. Lucha, and F. F. Sch\"oberl, Relativistic $N$-boson systems bound by pair potentials $V(r_{ij}) = g(r^2_{ij})$, J. Math. Phys. \textbf{45}, 3086 (2004).
\bibitem{silv10} B. Silvestre-Brac, C. Semay, F. Buisseret, and F. Brau, The quantum ${\cal N}$-body problem and the auxiliary field method, J. Math. Phys. \textbf{51}, 032104 (2010).
\bibitem{sema13a} C. Semay and C. Roland, Approximate solutions for $N$-body Hamiltonians with identical particles in $D$ dimensions, Res. in Phys. \textbf{3}, 231 (2013).
\bibitem{sema15a} C. Semay, Numerical Tests of the Envelope Theory for Few-Boson Systems, Few-Body Syst \textbf{56}, 149 (2015).
\bibitem{horn14} J. Horne, J. A. Salas, and K. Varga, Energy and Structure of Few-Body Systems, Few-Body Syst. \textbf{55}, 1245 (2014).
\bibitem{silv11} B. Silvestre-Brac and C. Semay, Duality relations in the auxiliary field method, J. Math. Phys. \textbf{52}, 052107 (2011).
\bibitem{sema15c} C. Semay, The Hellmann-Feynman theorem, the comparison theorem, and the envelope theory, Res. in Phys. \textbf{5}, 322 (2015).
\bibitem{sema15b} C. Semay, Improvement of the envelope theory with the dominantly orbital state method, Eur. Phys. J. Plus \textbf{130}, 156 (2015).
\bibitem{zimm75} H. E. Zimmerman, \textit{Quantum mechanics for organic chemists} (Academic Press, New York, 1975).
\bibitem{iwas03} M. Iwasaki, S.-I. Nawa, T. Sanada, and F. Takagi, Flux tube model for glueballs, Phys. Rev. D \textbf{68}, 074007 (2003).
\bibitem{meye05} H. B. Meyer and M. J. Teper, Glueball Regge trajectories and the pomeron: a lattice study, Phys. Lett. B \textbf{605}, 344 (2005).
\bibitem{buis09} F. Buisseret, V. Mathieu, and C. Semay, Glueball phenomenology and the relativistic flux tube model, Phys. Rev. D \textbf{80}, 074021 (2009).
\bibitem{yane94} R. J. Y\'a\~nez, W. Van Assche, and J. S. Dehesa, Position and momentum information entropies of the $D$-dimensional harmonic oscillator and hydrogen atom, Phys. Rev. A \textbf{50}, 3065 (1994).
\bibitem{gers10} V. D. Gershun and D. J. Cirilo-Lombardo, Higher spin particles in the discrete string model approach, J. Phys. A \textbf{43}, 305401 (2010).
\bibitem{lich89} D. B. Lichtenberg, Application of a generalized Feynman-Hellmann theorem to bound-state energy levels, Phys. Rev. D \textbf{40}, 4196 (1989).
\bibitem{luch90} W. Lucha, Relativistic Virial Theorems, Mod. Phys. Lett. A \textbf{5}, 2473 (1990).
\bibitem{silv12} B. Silvestre-Brac, C. Semay, and F. Buisseret, The Auxiliary Field Method in Quantum Mechanics, J. Phys. Math. \textbf{4}, P120601 (2012).
\bibitem{shur06} J. Liao and E. V. Shuryak, Polymer chains and baryons in a strongly coupled quark-gluon plasma, Nucl.\ Phys.\ A {\bf 775}, 224 (2006).

\end{thebibliography}
\end{document}